\documentclass[a4paper,11pt]{article}
\usepackage{pos}

\title{Improved precision in inverse beta decay cross section}
\author*[a,b]{Giulia Ricciardi}
\author[c,d]{Natascia Vignaroli}
\author[e]{Francesco Vissani}

\affiliation[a]{Departimento di Fisica E. Pancini, 
Universit\`a di Napoli Federico II,\\
Complesso Universitario di Monte Sant'Angelo, Via Cinthia, Napoli (NA), Italy }
\affiliation[b]{INFN, Sezione di Napoli,\\
Complesso Universitario di Monte Sant'Angelo, Via Cinthia, Napoli (NA), Italy }

\affiliation[c]{Dipartimento di Matematica e Fisica E. De Giorgi, Universit\`a del Salento, 73100 Lecce (LE), Italy }
\affiliation[d]{INFN, Sezione di Lecce, 73100 Lecce (LE), Italy}
\affiliation[e]{INFN, Laboratori Nazionali del Gran Sasso,\\67100 Assergi, L’Aquila (AQ), Italy.}

\emailAdd{giulia.ricciardi2@iunina.it}
\emailAdd{natascia.vignaroli@le.infn.it}
\emailAdd{francesco.vissani@lngs.infn.it}

\abstract{We analyze the cross section for inverse beta decay, focusing on the moderate energies (a few MeV to hundreds of MeV) relevant for reactor and supernova neutrinos. We discuss the updated evaluations of values and uncertainties in the cross section, and  the effect of  second-class currents. The estimate of theoretical precision is important for current and future experiments, when large data samples are or become available.}

\FullConference{
}

\usepackage{xcolor}

\begin{document}
\maketitle

\section{Introduction}

The cross section of the  inverse beta decay (IBD) is by far the most important mechanism of interaction of low-energy antineutrinos at relatively 
low energies for detectors that are based on water or hydrocarbons, i.e. the most commonly
used detectors such as scintillators or Cherenkov light detectors.

We mainly follow  Ref.  \cite{Ricciardi:2022pru}, which updates previous estimates of IBD cross section provided more than 20 years ago \cite{Vogel:1999zy,Strumia:2003zx}. The update considers recent experimental advances, improvements in reaction parameters, and the role of second-class currents (SCCs), which were previously omitted. We emphasize the importance of refining both the cross-section value and its uncertainty in an energy ranges about 3-70 MeV. 
That concerns the detection of reactor antineutrinos, e.g. at Daya Bay, whose full data sample comtains $ 5.55 \times 10^6$ reactor $\bar \nu_e$ candidates identified
as IBD interactions, and at the future detector JUNO, which  expects to
collect 83 events/day.  That also concerns studies of 
neutrinos from supernovae,  e.g. at 
Super-Kamiokande and JUNO, which anticipate to  collect more than
5000 IBD events for a typical galactic distance of
10kpc. An higher statistics is expected from the future Hyper-Kamiokande detector, which will have a mass about 10 times greater than Super-Kamiokande. 
The conservative 0.4\% 
 uncertainty estimate found in older papers \cite{Strumia:2003zx} turns out to be quite significant for large statistical samples, calling for a reassessment of the estimate. 
 
Within these energy ranges, we identify \cite{Ricciardi:2022pru} the primary uncertainties for the required level of accuracy, which are due to:
\begin{itemize} \item

     the Cabibbo angle and the axial coupling (at lower energies);
 \item  the vector axial radius $r_A$ (at higher energies).
 \end{itemize}
 In the following, we describe the key points of the theoretical cross section calculation and discuss the associated uncertainties. 
\section{Inverse beta decay cross section}

  One possible formulation of the most general matrix element of the charged weak current between proton and neutron states is
\begin{align}
\begin{split}
\mathcal{J}_\mu = \,
&\bar{u}_n \bigg(  f_1 \gamma_\mu +  g_1 \gamma_\mu \gamma_5 + i f_2 \sigma_{\mu\nu}\frac{q^\nu}{2M} + g_2  \frac{q_\mu}{M} \gamma_5  + f_3  \frac{q_\mu}{M} +  i g_3 \sigma_{\mu\nu}\frac{q^\nu}{2M} \gamma_5   \bigg) u_p
\end{split}
\label{eq:current}
\end{align}
This current includes all possible independent terms  which can be constructed from the Dirac $\gamma$ matrices and the hadronic quadrivectors $p_p$ and $p_n$.
The six scalar form factors depend upon the 
four-momentum transfer squared $t=q^2=-Q^2$, where  $q=p_\nu-p_e=p_n-p_p$. 
%They may depend on both the neutron $m_n$ and proton $m_p$ masses,  or --equivalently-- on the average nucleon mass $M =(m_n+m_p)/2$ and the nucleon mass difference  $\Delta = m_n -m_p$. 
%
The form factors $f_1$, $f_2$ and  $f_2$  are generally referred to, respectively, as vector, weak magnetism and scalar. 
The terms including them represent the vector part of the current. The terms including $g_1$, $g_2$ and  $g_2$  represent the axial part of the current.
The vector and axial-vector currents of the SM with form factors $f_1$, $f_2$, $g_1$ and $g_2$  are first class currents, the vector and axial vector currents with form factors $f_3$ and $g_3$ are second class currents.

We have calculated the cross section for the inverse beta decay, that is the process
\begin{equation}
\bar\nu_{\mbox{\tiny e}} (p_\nu) + \mbox{p}  (p_p) \to  \mbox{e}^+(p_e) + \mbox{n}(p_n)
\end{equation}
%This charged-current quasielastic (CCQE) interaction is the dominant reactions of electron antineutrinos till 2 GeV;
%for the reasons explained in the introduction, 
%we will be especially interested in the low energy part of this region. 
We have defined
\begin{equation}
\Delta = m_n -m_p \approx 1.293 \, \text{MeV}  \qquad M =\frac{m_n+m_p}{2} \approx  938.9 \, \text{MeV}
\end{equation}
The differential cross section is given by
\begin{equation}
\frac{d\sigma}{dt} =  \frac{G_F^2 \cos^2\theta_C }{64 \pi (s-m_p^2)^2}   \,  \overline{ |\mathcal{M}^2|}
\end{equation}
where $G_F$ is the Fermi coupling, $\theta_C$ the Cabibbo angle
(that is linked to the $u-d$ element of the CKM matrix by $ \cos\theta_C=V_{\mbox{\tiny ud}}$) and the  matrix element is:
\begin{align}
\begin{split}
\mathcal{M} = &\, \bar{v}_\nu \gamma^a(1-\gamma_5) v_e \cdot \\
&\bar{u}_n \bigg(  f_1 \gamma_a +  g_1 \gamma_a \gamma_5 + i f_2 \sigma_{ab}\frac{q^b}{2M} + g_2  \frac{q_a}{M} \gamma_5  + f_3  \frac{q_a}{M} +  i g_3 \sigma_{ab}\frac{q^b}{2M} \gamma_5   \bigg) u_p.
\end{split}
\end{align}
A straightforward calculation gives
\begin{equation}
 \overline{ |\mathcal{M}^2|} = A_{\bar \nu}(t) - (s-u)  B_{\bar \nu}(t) + (s-u)^2 C_{\bar \nu}(t) 
\end{equation}
where $s = (p_\nu + p_p)^2$, $t =q^2= (p_\nu - p_e)^2<0$, 
$u = (p_\nu - p_n)^2$ are the usual Mandelstam variables. %The calculation of the cross section is straightforward.
The  analytical formulas  of the functions  $A_{\bar \nu}(t)$, $B_{\bar \nu}(t)$, $C_{\bar \nu}(t)$, which include  second class currents, are given in Ref. \cite{Ricciardi:2022pru}.

\section{Uncertainties} The radiative corrections in QED to the cross section are calculated at leading order and included. Corrections at the next order and other effects, such as isospin breaking, are estimated to be small.
Hence, the primary uncertainties arise from  input parameters, whose significance is different at lower or higher energies \cite{Ricciardi:2022pru}:\begin{enumerate}
    \item 
    The main source for uncertainty on the IBD cross section under a few tens of MeVs
is due to the uncertainties on two input constants: the value of the CKM matrix element
 $V _ {\mbox{\tiny ud}}$. that multiplies the amplitude of transition,  and that of the axial coupling $g_1 (0) = \lim_{q^2 \to 0} g_1 (q^2)$ in Eq. \eqref{eq:current}.\item 
  At higher energies,  on the other hand, 
   the uncertainty depends 
also on the behavior of the form factors as $Q^2$ varies and thus on the parameters that describe such behavior, such as the axial mass $M_A$, or more precisely, the axial radius $r_A$.
\end{enumerate}
We discuss these two regions separately.

\subsection{The lower energy region}
The direct way to extract $V_{\mbox{\tiny ud}}=\cos \theta_C$
is to use the measurements of the super-allowed (s.a.) $0^+ \to 0^+$ charged current transitions. Being pure vector transitions, they only depend  on the better understood vector form factors. From 2006 to 2020, Hardy and Towner produced analyses of all decays this way,  with the more  recent estimate  \cite{Hardy:2020qwl}
$|V_{\mbox{\tiny ud}}{\mbox{\small (s.a.)}}| = 0.9737 (3) $, 
which we have adopted in Ref. \cite{Ricciardi:2022pru}. The extraction of $|V_{\mbox{\tiny ud}}{\mbox{\small (s.a.)}}| $ depends on the so-called
inner or universal electroweak radiative corrections (RC) to superallowed nuclear beta decays. These have been recently calculated  by a 
dispersion relation (DR) calculational approach \cite{Seng:2018yzq}. An average including the DR result has been provided in Ref. \cite{Cirigliano:2022yyo} and give a more precise value for the RC corrections, which leads to
$|V_{\mbox{\tiny ud}}{\mbox{\small (s.a.)}}| = 0.97367(11)_{exp}(13)_{RC}(27)_{NS}$, where $RC$ and $NS$ are the uncertainties associated to the radiative corrections and nuclear structure, respectively.
The first lattice QCD calculation of the RC  contributions yields
$|V_{\mbox{\tiny ud}}{\mbox{\small (s.a.)}}| =0.97386 (11)_{exp}(9)_{RC}(27)_{NS}$ \cite{Ma:2023kfr}.
The two have the same the total error, (32), and are consistent. These new results confirm the reasoning leading to choice made in Ref. \cite{Ricciardi:2022pru}.

%Using this vale  and combining it with$|V_{\mbox{\tiny us}}| = 0.2245 (8) $
%the tension with CKM unitarity is reduced to 1.8 $\sigma$.

An indirect determination of $|V_{\mbox{\tiny us}}| $ follows from noting that,
in the context of the SM, the CKM matrix is unitary. Using   the value  $|V_{\mbox{\tiny ud}}{\mbox{\small (s.a.)}}|$ along with  two determinations of
$|V_{\mbox{\tiny us}}| = 0.22431 (85)$  with S=2.5 \cite{ParticleDataGroup:2024cfk} and
  $ |V_{\mbox{\tiny ub}}| = 3.84 (26) \times 10^{- 3} $ given by averaging inclusive and exclusive decays \cite{Belle:2023asx}, which is consistent with CKMfitter $ |V_{\mbox{\tiny ub}}| = 3.64 (7) \times 10^{- 3} $  at 0.8$\sigma$,   the first row unitarity requirement seems to be violated. According to which experiment to measure the CKM parameters and inputs, this discrepancy goes from (about) 1 to 3$\sigma$.
 %we get
%V_{\mbox{\tiny ud}}{\mbox{\small (unit)}} = 0.9745 (2) $.
 This has fueled speculation regarding possible BSM models that could be driving deviations of the CKM matrix from unitarity. However,  considering the statistical consistency of this relatively weak inference and the success of the SM at the relevant energies, we believe it is reasonable to assume that this anomaly simply indicates the limits of current interpretations and measurements.

The axial coupling is often expressed as the ratio
$ \lambda = - g_1 (0)/f_1 (0)$, which is possible to measure  directly. The  average value is 
$ \lambda = 1.2754(13) $\cite{ParticleDataGroup:2024cfk},  which is not far from the lattice  average  in Ref. \cite{FlavourLatticeAveragingGroupFLAG:2024oxs}. The  parameters $\lambda$ and $V_{\mbox{\tiny ud}}$ are linked to the average lifetime of the neutron $\tau_{\mbox{\tiny n}}$ by the 
theoretical prediction \cite{Czarnecki:2019mwq}
\begin{equation}
\frac{1}{\tau_{\mbox{\tiny n}}} = \frac{V_{\mbox{\tiny ud}} ^ 2 \ (1+ 3 \lambda ^ 2)} {4906.4 \pm 1.7 \mbox{s}} \, ,
\label{eq:constraints}
\end{equation}
which holds since the average lifetime of the neutron and the IBD cross section depends on the same matrix element. 
By propagating the errors, including RC, one can  find a prediction for $\tau_{\mbox{\tiny n}}$.
%$ \tau_{\mbox{\tiny n}}{\mbox{\small(SM)}} = 878.38 \pm 0.89 $~s 
% dipende
%dal valore di lambda. Inoltre cambiare le RC equivale al primo ordine a  moltiplicare a dx per ( 1+RC). 
%By measuring  $ \tau_{\mbox{\tiny n}} $, one obtains a constraint on  $\lambda$ and $V_{\mbox{\tiny ud}}$.
Vice versa, by measuring $\tau_{\mbox{\tiny n}}$ it is possible to constrain  $\lambda$ and $V_{\mbox{\tiny ud}}$.

The neutron decay lifetime plays a key role also in cosmology. The neutron/proton ratio during nucleosynthesis
depends strongly on the neutron lifetime and
 its uncertainty stands in the way of precise predictions in Standard Big Bang Nucleosynthesis. 
There are two methods to measure $ \tau_{\mbox{\tiny n}} $, usually indicated as the bottle  and the beam method. In the former, ultra-cold neutrons (UCN) are trapped and their number is measured over time, yielding $ \tau^{bottle}_{\mbox{\tiny n}}= 878.4 \pm 0.5 $
 s \cite{{ParticleDataGroup:2024cfk}}. 
 In the beam method, 
neutron  decay products are counted relative to the number of incident neutrons, determining the  average lifetime 
of $ \tau^{beam}_{\mbox{\tiny n}}= 888.0 \pm 2.0 $
s. The  discrepancy between the two methods is known as
the neutron lifetime puzzle, raising concerns about
the reliability of neutron lifetime measurements.  The value predicted by Eq. \eqref{eq:constraints} is closer to the lower value, which has prompt Ref. \cite{Ricciardi:2022pru} to use only data obtained by the bottle method. The assumption is that the dataset from beam experiment is affected by a systematic deviation, 
 not yet fully understood.
Recently, an experiment has been performed with the beam method but differing from previous experiments that measured protons. Instead, it 
detected electrons, enabling measurements with distinct systematic uncertainties. The result is 
$\tau^{beam}_{\mbox{\tiny n}} = 877.2 \pm {1.7_{stat}}^{+ 4.0}_{- 3.6_{stat}} $~s \cite{Fuwa:2024cdf}.
This value is consistent with bottle method measurements but exhibits a 2.3$\sigma$ tension with the
average value obtained from the proton-detection-based beam method.

\subsection{The higher energy region}
There is not an unique way to express the dependence on $t$ of the form factors. A rather common procedure is to adopt {\em phenomenological} descriptions of the behaviour of the form factors of the nucleons~\cite{LlewellynSmith:1971uhs}. It should be noted that the phenomenological form factors, especially the dipolar approximation, are not optimized at low energies, and in some cases, the differences can be significant.
Alternatively, the form factors  can be constrained  by using analytic methods, crossing symmetry and  global fits, which include several intermediate states and continuum contributions. Finally, the form factors can be calculated {\em ab initio}, in particular, by exploiting  lattice QCD.

Regardless of the full form factor dependence on $t$, when the relevant energies are relatively low, 
 we can confidently assert that  only the first terms of its Taylor expansion are needed to assess its uncertainty.
In the case of supernova neutrino detection, for instance, we have $Q^2_{\mbox{\tiny max}}\sim (2 E)^2\lesssim 0.01\mbox{ GeV}^2$ ($Q^2=-t>0$)
and the higher order terms in $Q^2$ can be safely neglected. 
Hence we will use the form factors in their linearized form.
%
%%%%%https://arxiv.org/pdf/1403.2673.pdf formula 23
%
The axial vector form factor $g_1(t)$  is  the one that causes the largest uncertainty.
Following conventional usage,
  we  
define  a vector axial radius
 $\sqrt{\langle r_{\mbox{\tiny\rm A}}^2 \rangle}$ where
\begin{equation}
\langle r_{\mbox{\tiny\rm A}}^2 \rangle = \frac{6}{g_1(0)} \left.  \frac{d g_1(t)}{dt^2} \right|_{t=0}
\end{equation}
The linear expansions 
in terms of the vector axial radius becomes
\begin{equation}
    \frac{ g_1(t)}{g_1(0)}   
\equiv 
 1 - \frac{\langle r_{\mbox{\tiny\rm A}}^2 \rangle \ Q^2}{6}  + \mathcal{O}(Q^4). 
\end{equation}
In this framework, we can define  
\begin{equation}
M_{\mbox{\tiny\rm A}}^2 \equiv -2 \frac{{g_1}'(0) }{g_1(0)} = 
\frac{12}{\langle r_{\mbox{\tiny\rm A}}^2 \rangle }
\end{equation}
which can be used to compare with the description adopting the traditional  dipole parametrization.
For the energies of interest for the detection of supernova neutrinos, we have $Q^2_{\rm max}\sim (2 E)^2\lesssim  0.01\mbox{ GeV}^2$  for  $E<50$ MeV.
T%hus, using $M_{\mbox{\tiny\rm A}}^2\sim \mbox{GeV}^2$, 
%we find also in this case that  $g_1(Q^2)$ varies by approximately $2$\% and higher order terms in $Q^2$ have a negligible role.

The vector axial 
radius is traditionally probed using:
1) direct measurements of charged current interactions, in particular 
muon neutrinos on Deuterium targets;
2) 
 muon capture on proton
3) single pion production by 
electrons on nucleons.

In $\nu\mbox{N}$ 
direct measurements  using 
muon neutrinos on Deuterium, the formal errors  are  
very small, giving $r_{\mbox{\tiny\rm A}}^2 =0.453\pm 0.023$ fm$^2$ \cite{Hill:2017wgb, Bodek:2007ym},
These measurements are obtained at higher energies than those we are interested in, and therefore cannot be used 
without an extrapolation. According to \cite{Bhattacharya:2011ah} this extrapolation is not without dangers, hence  
Ref.~\cite{Hill:2017wgb} suggests to use $r_{\mbox{\tiny\rm A}}^2 =0.46\pm 0.22$ fm$^2$, which does not rely on the dipole approximation
and is consistent with the previous results, but 
has a considerably larger error.  

The muon capture on proton, $\mu \,+\, p \to \nu_\mu \,+\,  n$, due to crossing invariance, 
 probes the exact form factors at small $Q^2$ we are interested in, making it highly relevant to our case.
 The  MuCap (Muon Capture on the Proton) experiment 
has collected data from 2004–2007 at the Paul Scherrer
Institute (PSI) in Switzerland using a time projection
chamber  to detect low-energy
muons. It has given results that are quite accurate and consistent with those
of the previous method, with errors of similar magnitude \cite{Hill:2017wgb}.

Experiments on single pion production by 
electrons on nucleons give extremely precise results, 
and values consistent with previous ones \cite{Hill:2017wgb, Bodek:2007ym}, that - formally - cover precisely the most interesting $Q^2$ region, see e.g.~\cite{Bernard:2001rs}. 
Unfortunately they require using the theory in a regime where its reliability, according to \cite{Hill:2017wgb},  can be doubted. 

Recently, the Miner$\nu$a  collaboration has 
 measured the axial vector form factor from antineutrino–proton scattering \cite{MINERvA:2023avz}. A muon antineutrino elastically scatters off the free proton
from the hydrogen atom, turning the neutrino into the 
positively charged muon  and the proton into a neutron. The reaction $\bar \nu_\mu + p \to \mu^+ + n $ is free from 
nuclear theory corrections in scattering from deuterium  and provides a direct measurement of $g_1$. It is also a
two-body reaction with a nucleon at rest; therefore, the neutrino
direction and the final-state muon  momentum fully specify the interacting system. The beam produced at the NuMI neutrino beamline  at Fermilab
has an average energy of 5.4 GeV. The resulting value is 
$r_{\mbox{\tiny\rm A}}=0.73\pm 0.17$ fm.
%It reaches an higher precision, while being  compatible with average values from $\nu\mbox{N}$. 

\section{Conclusions}

 Neutrinos of varying energies are detected in different experiments. We focus on  the energy range that on the low side encompasses reactor neutrinos (up to approximately 10 MeV) and on  the higher side neutrino fluxes from supernovae (up to 50 MeV).
 
We have discussed the theoretical  uncertainty of the IBD cross section, which
%We hSecond-class currents are not expected to make a significant contribution. 
 depends critically upon the set of three parameters $V _ {\mbox{\tiny ud}}$, $\lambda$ and  the vector axial radius $r_A$.  Reliable measurements of key parameters are essential to meet the precision advocated by present and future experiments. 
Theoretical advances in lattice QCD are anticipated, which will lead to   estimates with uncertainties smaller than the present ones. Meanwhile, significant experimental progress in the determination and analysis of these parameters has been made in recent years.

\section*{Aknowledgments}

G.R. thanks the Organizing Committee of the XVIth Quark Confinement and Hadron Spectrum Conference ("Confinement24", Cairns, Australia, 19-24 August, 2024) for the kind invitation.
This work is partially supported by the INFN research initiative ENP.


\begin{thebibliography}{99}
%\cite{Ricciardi:2022pru}
\bibitem{Ricciardi:2022pru}
G.~Ricciardi, N.~Vignaroli and F.~Vissani,
%``An accurate evaluation of electron (anti-)neutrino scattering on nucleons,''
JHEP \textbf{08} (2022), 212
doi:10.1007/JHEP08(2022)212
[arXiv:2206.05567 [hep-ph]].
\bibitem{Vogel:1999zy}
P.~Vogel and J.~F.~Beacom,
%``Angular distribution of neutron inverse beta decay, anti-neutrino(e) + p ---\ensuremath{>} e+ + n,''
Phys. Rev. D \textbf{60} (1999), 053003
doi:10.1103/PhysRevD.60.053003
[arXiv:hep-ph/9903554 [hep-ph]].
%636 citations counted in INSPIRE as of 01 Jun 2022
\bibitem{Strumia:2003zx}
A.~Strumia and F.~Vissani,
%``Precise quasielastic neutrino/nucleon cross-section,''
Phys. Lett. B \textbf{564} (2003), 42-54
doi:10.1016/S0370-2693(03)00616-6
[arXiv:astro-ph/0302055 [astro-ph]].
\bibitem{Hardy:2020qwl}
J.~C.~Hardy and I.~S.~Towner,
%``Superallowed $0^+ \to 0^+$ nuclear $\beta$ decays: 2020 critical survey, with implications for V$_{ud}$ and CKM unitarity,''
Phys. Rev. C \textbf{102} (2020) no.4, 045501
doi:10.1103/PhysRevC.102.045501
%227 citations counted in INSPIRE as of 19 Feb 2025
%\cite{Seng:2018yzq}
\bibitem{Seng:2018yzq}
C.~Y.~Seng, M.~Gorchtein, H.~H.~Patel and M.~J.~Ramsey-Musolf,
%``Reduced Hadronic Uncertainty in the Determination of $V_{ud}$,''
Phys. Rev. Lett. \textbf{121} (2018) no.24, 241804
doi:10.1103/PhysRevLett.121.241804
[arXiv:1807.10197 [hep-ph]].
%\cite{Cirigliano:2022yyo}
\bibitem{Cirigliano:2022yyo}
V.~Cirigliano, A.~Crivellin, M.~Hoferichter and M.~Moulson,
%``Scrutinizing CKM unitarity with a new measurement of the K\ensuremath{\mu}3/K\ensuremath{\mu}2 branching fraction,''
Phys. Lett. B \textbf{838} (2023), 137748
doi:10.1016/j.physletb.2023.137748
[arXiv:2208.11707 [hep-ph]].
%\cite{Ma:2023kfr}
\bibitem{Ma:2023kfr}
P.~X.~Ma, X.~Feng, M.~Gorchtein, L.~C.~Jin, K.~F.~Liu, C.~Y.~Seng, B.~G.~Wang and Z.~L.~Zhang,
%``Lattice QCD Calculation of Electroweak Box Contributions to Superallowed Nuclear and Neutron Beta Decays,''
Phys. Rev. Lett. \textbf{132} (2024) no.19, 191901
doi:10.1103/PhysRevLett.132.191901
[arXiv:2308.16755 [hep-lat]].
\bibitem{ParticleDataGroup:2024cfk}
S.~Navas \textit{et al.} [Particle Data Group],
%``Review of particle physics,''
Phys. Rev. D \textbf{110} (2024) no.3, 030001
doi:10.1103/PhysRevD.110.030001
%997 citations counted in INSPIRE as of 19 Feb 2025
\bibitem{Belle:2023asx}
L.~Cao \textit{et al.} [Belle],
%``First Simultaneous Determination of Inclusive and Exclusive |Vub|,''
Phys. Rev. Lett. \textbf{131} (2023) no.21, 211801
doi:10.1103/PhysRevLett.131.211801
[arXiv:2303.17309 [hep-ex]].
\bibitem{FlavourLatticeAveragingGroupFLAG:2024oxs}
Y.~Aoki \textit{et al.} [Flavour Lattice Averaging Group (FLAG)],
%``FLAG Review 2024,''
[arXiv:2411.04268 [hep-lat]].
%38 citations counted in INSPIRE as of 19 Feb 2025
\bibitem{Czarnecki:2019mwq}
A.~Czarnecki, W.~J.~Marciano and A.~Sirlin,
``Radiative Corrections to Neutron and Nuclear Beta Decays Revisited,''
Phys. Rev. D \textbf{100} (2019) no.7, 073008
\bibitem{UCNt:2021pcg}
F.~M.~Gonzalez \textit{et al.} [UCN\ensuremath{\tau}],
Phys. Rev. Lett. \textbf{127} (2021) no.16, 162501
doi:10.1103/PhysRevLett.127.162501
[arXiv:2106.10375 [nucl-ex]].
\bibitem{Fuwa:2024cdf}
Y.~Fuwa, T.~Hasegawa, K.~Hirota, T.~Hoshino, R.~Hosokawa, G.~Ichikawa, S.~Ieki, T.~Ino, Y.~Iwashita and M.~Kitaguchi, \textit{et al.}
\bibitem{Hill:2017wgb}
R.~J.~Hill, P.~Kammel, W.~J.~Marciano and A.~Sirlin,
%``Nucleon Axial Radius and Muonic Hydrogen \textemdash{} A New Analysis and Review,''
Rept. Prog. Phys. \textbf{81} (2018) no.9, 096301
doi:10.1088/1361-6633/aac190
[arXiv:1708.08462 [hep-ph]].
%``Improved measurements of neutron lifetime with cold neutron beam at J-PARC,''
[arXiv:2412.19519 [nucl-ex]].
%\cite{LlewellynSmith:1971uhs}
\bibitem{LlewellynSmith:1971uhs}
C.~H.~Llewellyn Smith,
%``Neutrino Reactions at Accelerator Energies,''
Phys. Rept. \textbf{3} (1972), 261-379
\bibitem{Bodek:2007ym}
A.~Bodek, S.~Avvakumov, R.~Bradford and H.~S.~Budd,
%``Vector and Axial Nucleon Form Factors:A Duality Constrained Parameterization,''
Eur. Phys. J. C \textbf{53} (2008), 349-354
doi:10.1140/epjc/s10052-007-0491-4
[arXiv:0708.1946 [hep-ex]].
\bibitem{Bhattacharya:2011ah}
B.~Bhattacharya, R.~J.~Hill and G.~Paz,
%``Model independent determination of the axial mass parameter in quasielastic neutrino-nucleon scattering,''
Phys. Rev. D \textbf{84} (2011), 073006
doi:10.1103/PhysRevD.84.073006
[arXiv:1108.0423 [hep-ph]].
\bibitem{Bernard:2001rs}
V.~Bernard, L.~Elouadrhiri and U.~G.~Meissner,
%``Axial structure of the nucleon: Topical Review,''
J. Phys. G \textbf{28} (2002), R1-R35
doi:10.1088/0954-3899/28/1/201
[arXiv:hep-ph/0107088 [hep-ph]].
\bibitem{MINERvA:2023avz}
T.~Cai \textit{et al.} [MINERvA],
%``Measurement of the axial vector form factor from antineutrino\textendash{}proton scattering,''
Nature \textbf{614} (2023) no.7946, 48-53
doi:10.1038/s41586-022-05478-3
\end{thebibliography}
\end{document}